\documentclass[a4paper]{amsart}

\usepackage{geometry}
\usepackage{graphicx}
\usepackage[font=scriptsize]{caption}
\usepackage{tikz}
\usetikzlibrary{trees,positioning}

\usepackage[applemac]{inputenc}
\usepackage[T1]{fontenc}

\RequirePackage{amsmath}
\RequirePackage{bm}
\RequirePackage{amssymb}
\RequirePackage{upref}
\RequirePackage{amsthm}
\RequirePackage{enumerate}
\RequirePackage{pb-diagram}
\RequirePackage{amsfonts}
\RequirePackage[mathscr]{eucal}
\RequirePackage{verbatim}
\RequirePackage{xr}
\RequirePackage{graphicx}
\usepackage{calc}
\usepackage{xspace}
\RequirePackage{color}
\RequirePackage{ifthen}
\RequirePackage{fancybox}
\RequirePackage{mathrsfs}

\newcommand{\cf}{cf.\@\xspace}

\newcommand{\al}{\alpha}
\newcommand{\bet}{\beta}
\newcommand{\ga}{\gamma}
\newcommand{\de}{\delta }

\newcommand{\f}{\varphi}
\newcommand{\h}{\eta}

\newcommand{\ka}{\kappa}
\newcommand{\lam}{\lambda}
\newcommand{\m}{\mu}

\newcommand{\vt}{\vartheta}

\newcommand{\s}{\sigma}

\newcommand{\C}{\varGamma}
\newcommand{\D}{\varDelta}
\newcommand{\F}{\varPhi}
\newcommand{\Lam}{\varLambda}
\newcommand{\Om}{\varOmega}

\newcommand{\so}{{\mc S_0}}
\newcommand{\socc}{{\mc S_0}}

\newcommand{\const}{\tup{const}}

\newcommand{\msp[1]}[1]{\mspace{#1mu}}

\newcommand{\R}[1][n+1]{{\protect\mathbb R}^{#1}}

\newcommand{\Ss}[1][n+1]{{\protect\mathbb S}^{#1}}

\newcommand{\N}{{\protect\mathbb N}}

\newcommand{\Z}{{\protect\mathbb Z}}
\newcommand{\eR}{\stackrel{\lower1ex \hbox{\rule{6.5pt}{0.5pt}}}{\msp[3]\R[]}}
\newcommand{\eN}{\stackrel{\lower1ex \hbox{\rule{6.5pt}{0.5pt}}}{\msp[1]\N}}
\newcommand{\eO}{\stackrel{\lower1ex \hbox{\rule{6pt}{0.5pt}}}{\msc O}}
\newcommand{\mf}[1]{\mathfrak {#1}}

\DeclareMathOperator{\imc}{Im}

\DeclareMathOperator{\grad}{grad}

\DeclareMathOperator{\tr}{tr}

\DeclareMathOperator{\diag}{diag}

\newcommand\ra{\rightarrow}

\newcommand\pa{\partial}
\newcommand\pde[2]{\frac {\partial#1}{\partial#2}}
\newcommand\pd[3]{\frac {\partial#1}{\partial#2^#3}}   
 
\newcommand{\un}{\infty}
\newcommand{\A}{\forall}

\newcommand{\uu}{\cup}
\newcommand{\ii}{\cap}
\newcommand{\uuu}{\bigcup}

\newcommand{\uud}{ \stackrel{\lower 1ex \hbox {.}}{\uu}}
\newcommand{\uuud}[1]{ \stackrel{\lower 1ex \hbox {.}}{\uuu_{#1}}}
\newcommand\su{\subset}

\newcommand{\sminus}[1][28]{\raise 0.#1ex\hbox{$\scriptstyle\setminus$}}

\newcommand{\wh}{\widehat}

\newcommand{\wed}{\wedge}

\newcommand{\abs}[1]{\lvert#1\rvert}

\newcommand{\norm}[1]{\lVert#1\rVert}

\newcommand{\spd}[2]{\protect\langle #1,#2\protect\rangle}

\newcommand{\tit}{\textit}

\newcommand{\tup}{\textup}

\newcommand{\mc}{\protect\mathcal}
\newcommand{\msc}{\protect\mathscr}

\newcommand{\tlam}{\tilde\lam}

\providecommand{\bysame}{\makebox[3em]{\hrulefill}\thinspace}

\newcommand{\bt}{\begin{thm}}
\newcommand{\bl}{\begin{lem}}
\newcommand{\bc}{\begin{cor}}
\newcommand{\bd}{\begin{definition}}
\newcommand{\bpp}{\begin{prop}}
\newcommand{\br}{\begin{rem}}
\newcommand{\bn}{\begin{note}}
\newcommand{\be}{\begin{ex}}
\newcommand{\bes}{\begin{exs}}
\newcommand{\bb}{\begin{example}}
\newcommand{\bbs}{\begin{examples}}
\newcommand{\ba}{\begin{axiom}}
\newcommand{\bas}{\begin{assumption}}

\newcommand{\et}{\end{thm}}
\newcommand{\el}{\end{lem}}
\newcommand{\ec}{\end{cor}}
\newcommand{\ed}{\end{definition}}
\newcommand{\epp}{\end{prop}}
\newcommand{\er}{\end{rem}}
\newcommand{\en}{\end{note}}
\newcommand{\ee}{\end{ex}}
\newcommand{\ees}{\end{exs}}
\newcommand{\eb}{\end{example}}
\newcommand{\ebs}{\end{examples}}
\newcommand{\ea}{\end{axiom}}
\newcommand{\eas}{\end{assumption}}

\newcommand{\bp}{\begin{proof}}
\newcommand{\ep}{\end{proof}}
\newcommand{\eps}{\renewcommand{\qed}{}\end{proof}}

\newcommand{\bal}{\begin{align}}

\newcommand{\bi}[1][1.]{\begin{enumerate}[\upshape #1]}
\newcommand{\bia}[1][(1)]{\begin{enumerate}[\upshape #1]}
\newcommand{\bin}[1][1]{\begin{enumerate}[\upshape\bfseries #1]}
\newcommand{\bir}[1][(i)]{\begin{enumerate}[\upshape #1]}
\newcommand{\bic}[1][(i)]{\begin{enumerate}[\upshape\hspace{2\cma}#1]}
\newcommand{\bis}[2][1.]{\begin{enumerate}[\upshape\hspace{#2\parindent}#1]}
\newcommand{\ei}{\end{enumerate}}

\newcommand\ndots{\raise 0.47ex \hbox {,}\hskip0.06em\cdots %
     \raise 0.47ex \hbox {,}\hskip0.06em} 

\newcommand{\q}{\quad}
\newcommand{\qq}{\qquad}

\newcommand{\hp}{\hphantom}

\newcommand\nd{\noindent}

\newskip\Csmallskipamount                                                
\Csmallskipamount=\smallskipamount
\newskip\Cmedskipamount
\Cmedskipamount=\medskipamount
\newskip\Cbigskipamount
\Cbigskipamount=\bigskipamount

\newcommand\cvs{\vspace\Csmallskipamount}   
\newcommand\cvm{\vspace\Cmedskipamount}

\newskip\csa
\csa=\smallskipamount

\newskip\cma
\cma=\medskipamount

\newskip\cba
\cba=\bigskipamount

\newdimen\spt
\newcommand\citem{\cvs\advance\itemno by
1{(\romannumeral\the\itemno})\hskip3pt}
\newcommand{\bitem}{\cvm\nd\advance\itemno by
1{\bf\the\itemno}\hspace{\cma}}

\newcount\itemno
\newcommand{\las}[1]{\label{S:#1}}

\newcommand{\lae}[1]{\label{E:#1}}
\newcommand{\lat}[1]{\label{T:#1}}
\newcommand{\lal}[1]{\label{L:#1}}

\newcommand{\lac}[1]{\label{C:#1}}

\newcommand{\lar}[1]{\label{R:#1}}

\newcommand{\rt}[1]{Theorem~\ref{T:#1}}
\newcommand{\rl}[1]{Lemma~\ref{L:#1}}

\newcommand{\rc}[1]{Corollary~\ref{C:#1}}

\newcommand{\rr}[1]{Remark~\ref{R:#1}}

\newcommand{\re}[1]{\eqref{E:#1}}
\newcommand{\frc}[1]{Corollary~\ref{C:#1} on page~\tup{\pageref{C:#1}}}
\newcommand{\frt}[1]{Theorem~\ref{T:#1} on page~\tup{\pageref{T:#1}}}
\newcommand{\frl}[1]{Lemma~\ref{L:#1} on page~\tup{\pageref{L:#1}}}

\newcommand{\fre}[1]{\eqref{E:#1} on page~\tup{\pageref{E:#1}}}

\newskip\thmskip
\thmskip=\parindent

\newskip\hsk
\newenvironment{hinw}{\labelsep=0pt\begin{list}{}{\labelsep=0pt\itemindent=0pt\labelwidth=0pt\leftmargin=\parindent\rightmargin=0pt\partopsep=\cba}%
\item\it\nopagebreak\nopagebreak}%
{\end{list}}

\newcommand\bh{\begin{hinw}}
\newcommand{\eh}{\end{hinw}}

\newtheoremstyle{normal}
  {\cba}
  {\cba}
  {}
  {\thmskip}
  {\bfseries}
  {.}
  {\hsk}
  {}

\newtheoremstyle{abschnitt}
  {\cba}
  {\cba}
  {}
  {\thmskip}
  {\bfseries}
  {.}
  {\hsk}
  {}

\newtheoremstyle{italic}
  {\cba}
  {\cba}
  {\itshape}
  {\thmskip}
  {\bfseries}
  {.}
  {\hsk}
  {}

\newtheoremstyle{aufgaben}
  {\cba}
  {\cba}
  {}
  {}
  {\normalsize\bfseries}
  {.}
  {\hsk}
  {}

\newtheoremstyle{break}
  {\cba}
  {\cba}
  {\itshape}
  {}
  {\bfseries}
  {.}
  {\newline}
  {}

\theoremstyle{italic}
\newtheorem{thm}[subsection]{Theorem}
\newtheorem{lem}[subsection]{Lemma}
\newtheorem{prop}[subsection]{Proposition}
\newtheorem{cor}[subsection]{Corollary}

\theoremstyle{normal}
\newtheorem{rem}[subsection]{Remark}
\newtheorem{definition}[subsection]{Definition}
\newtheorem{example}[subsection]{Example}
\newtheorem{examples}[subsection]{Examples}
\newtheorem{ex}[subsection]{Exercise}
\newtheorem{note}[subsection]{}
\newtheorem{axiom}[subsection]{Axiom}
\newtheorem{assumption}[subsection]{Assumption}

\theoremstyle{aufgaben}
\newtheorem{exs}[subsection]{Exercises}

\numberwithin{equation}{section}
\numberwithin{figure}{section}

\newenvironment{textequation}[1][0.8]
{\begin{equation}
\begin{aligned}
\begin{minipage}{#1\linewidth}}
{\end{minipage}
\end{aligned}
\end{equation}
\ignorespacesafterend}

\newcommand{\btext}{\begin{textequation}}
\newcommand{\etext}{\end{textequation}}

\def\hinweis{\@startsection{subsection}{2}%
 \z@{0.7\linespacing\@plus 0.5\linespacing}{0.7\linespacing}%
{\normalfont\itshape\indent}}

\newcounter{hours}\newcounter{minutes}
\newcommand{\printtime}{%
\setcounter{hours}{\time/60}%
\setcounter{minutes}{\time-\value{hours}*60}%
\ifthenelse{\value{minutes}<10}{\thehours :0\theminutes}{\thehours:\theminutes}}

\usepackage[german,american]{babel}
\usepackage{graphicx}
\RequirePackage{amsmath}
\RequirePackage{bm}
\RequirePackage{amssymb}
\RequirePackage{upref}
\RequirePackage{amsthm}
\RequirePackage{enumerate}
\RequirePackage{pb-diagram}
\RequirePackage{amsfonts}
\RequirePackage[mathscr]{eucal}
\RequirePackage{verbatim}
\RequirePackage{xr}
\RequirePackage{graphicx}
\RequirePackage{mathrsfs}
\usepackage{calc}
\usepackage{xspace}
\usepackage{dsfont}
\newcommand{\Clk}{\varGamma}

\makeatletter
\RequirePackage{color}
\newcommand{\ann}[1]{\renewcommand{\@makefnmark}{\mbox{$^{\color{red}{\@thefnmark}}$}}%
\footnote {#1}}
\makeatother

\RequirePackage{upref}
\RequirePackage{amsthm}
\RequirePackage{enumerate}
\usepackage[mathscr]{eucal}

\usepackage{xr-hyper}

\usepackage{calc}

\newlength{\oddsidemarginlength}
\newlength{\topmarginlength}

\newcounter{numberoflines}
\newcounter{tempcc}
\oddsidemargin=\oddsidemarginlength
\evensidemargin=\oddsidemargin
\topmargin=\topmarginlength
\usepackage[colorlinks=true,linkcolor=blue,citecolor=blue,urlcolor=blue]{hyperref}  

\begin{document}

\flushbottom


\title[Quantizing the exterior region of a black hole]{Quantizing the exterior region of a Kerr-AdS black hole leads to a resolution of the information paradox at the quantum level}

\author{Claus Gerhardt}
\address{Ruprecht-Karls-Universit\"at, Institut f\"ur Angewandte Mathematik,
Im Neuenheimer Feld 205, 69120 Heidelberg, Germany}
\email{\href{mailto:gerhardt@math.uni-heidelberg.de}{gerhardt@math.uni-heidelberg.de}}
\urladdr{\href{http://www.math.uni-heidelberg.de/studinfo/gerhardt/}{http://www.math.uni-heidelberg.de/studinfo/gerhardt/}}

%


\date{\today}
%


\begin{abstract} 
We quantize the exterior region of an odd-dimensional Kerr-AdS black hole, in which all rotational parameters are equal, using our model of quantum gravity. The resulting hyperbolic equation is solved by products of temporal eigenfunctions $w_i$, whose corresponding eigenvalues all have multiplicity one, and spatial eigendistributions $v_{ij}$ corresponding to the same eigenvalues but having multiplicities $m_i\ge 1$. In principle, the $m_i$ may be arbitrarily large. When only the exterior region is considered, there is no criterion for determining the values of the $m_i$. However, when the quantization of the interior region is also considered, the same ambiguity does not arise, because the multiplicities can be chosen to be maximal. It is therefore natural to use the same values in the exterior region. Because the temporal Hamiltonian is the same in both regions, the interior and exterior eigenvalues coincide, and this choice defines a unitary equivalence between the respective Hilbert spaces and Hamiltonians. The unitary equivalence between the Hilbert spaces and their respective Hamiltonians preserves the complete quantum-statistical information carried by all normal states in the interior and exterior regions of the black hole. Hence, there is no information paradox at the quantum level.
\end{abstract}

\maketitle

\tableofcontents

\setcounter{section}{0}
\section{Introduction}

In \cite{cg:qbh25}, we quantized the \tit{exterior} region of a Schwarzschild-AdS black hole using our model for quantizing a globally hyperbolic spacetime $N=N^{n+1}$, $n\ge 3$, as described in \cite{cg:qgravity-book2}. The final result of the quantization process is a hyperbolic equation 
\begin{equation}\lae{1.1}
H_0{\tilde u}-H_1{\tilde u}=0
\end{equation}
in a quantum spacetime
\begin{equation}\lae{1.2}
Q=(0,\un)\times \so
\end{equation}
where $\so$ is a Cauchy hypersurface of $N$ with induced metric $\tilde g$. The temporal Hamiltonian $H_0$ is a self-adjoint operator with countably many eigenfunctions $w_i$ and corresponding eigenvalues $\lam_i>0$. The spatial Hamiltonian $H_1$ is also self-adjoint but has continuous spectrum.

To solve the hyperbolic equation, we must solve the eigenvalue equation 

\begin{equation}\lae{1.3}
H_1v_i=\lam_i v_i,
\end{equation}
for every $i\in\N$, where
\begin{equation}\lae{1.4}
H_1v=\frac {16(n-1)}n \{-(n-1)\D v-\frac{n-2}2 Rv\}
\end{equation}
and the Laplacian and scalar curvature refer to the induced metric on the Cauchy hypersurface of $N$. If $N$ is the exterior region of a black hole, then we choose
\begin{equation}
\so=\{t=\const\}.
\end{equation}
The solutions $\tilde u$ of \re{1.1} can then be expressed as products of the corresponding eigenfunctions or eigendistributions
\begin{equation}
\tilde u_i=w_iv_i\qq\A\,i\in \N.
\end{equation}
The temporal eigenvalues $\lam_i$ all have multiplicity one, whereas the spatial eigenvalues generally have multiplicity greater than one. Hence, we label the spatial eigenfunctions in the form
\begin{equation}
v_{ij},\qq 1\le j\le m_i,
\end{equation}
such that
\begin{equation}\lae{1.9}
H_1v_{ij}=\lam_i v_{ij}, \qq\A\,1\le j\le m_i.
\end{equation}
To apply quantum statistics to this configuration, one must define a Hilbert space in which the $v_{ij}$ form a complete orthonormal basis, prove that $H_1$ is self-adjoint and that
\begin{equation}
e^{-\bet H_1}
\end{equation}
is of trace class for $\bet>0$, and verify that the multiplicities $m_i$ satisfy appropriate physical or logical criteria. In particular, they cannot be chosen arbitrarily or be arbitrarily large. 

In the case of a Schwarzschild-AdS black hole, the multiplicities $m_i$ could be chosen arbitrarily large for any fixed $i$. Consequently, the von Neumann entropy and average energy could also be made arbitrarily large, rendering the quantum-statistical description useless. By contrast, in the quantization of the interior region of the Schwarzschild-AdS black hole, the multiplicities could be defined by maximizing their values. The corresponding Cauchy hypersurfaces were the spacelike slices $S_r$, $0<r<r_0$, where $r_0$ is the black hole radius. We could also let $r\ra r_0$, because the induced metrics of the slices converged to a smooth Riemannian metric, so that the spatial Hamiltonian $H_1$ and the multiplicities, denoted by $n(\lam_i)$, corresponded to the event horizon. Hence, the simplest and most logical choice for defining the multiplicities $m_i$ in the exterior region was to set
\begin{equation}\lae{1.10}
m_i=n(\lam_i)\qq\A\, i\in \N.
\end{equation}
Hence, the quantum-statistical results in both regions were identical. Furthermore, because the spatial eigendistributions in the two regions form orthonormal bases of their respective Hilbert spaces, relation \re{1.10} can be used to define a unitary map between these Hilbert spaces under which the respective spatial Hamiltonians are unitarily equivalent, since the corresponding eigenvalues coincide. Hence, the black hole information paradox does not exist at the quantum level.

In this paper, we quantize the exterior region of an odd-dimensional Kerr-AdS black hole $N=N^{2m+1}$, $m\ge 2$, in which all rotational parameters are equal. The problems encountered in the Schwarzschild-AdS case are also present when quantizing the exterior region of a rotating black hole. We shall show that only a combination of the results for the interior and exterior regions yields a satisfactory quantum-statistical description of the exterior region. In addition, we shall prove that the unitary equivalence between the Hilbert spaces and their respective Hamiltonians preserves the complete quantum-statistical information carried by all normal states in the interior and exterior regions of the black hole. Hence, the information paradox does not exist.

We shall also prove:
\bt
The spatial eigenfunctions $v_{ij}=v_{ij}(\tau,x)$ may be regarded as gravitational radiation modes that emanate from the event horizon and decay exponentially at infinity, satisfying 
\begin{equation}
v_{ij}(0,x)=0
\end{equation}
and
\begin{equation}
\lim_{\tau\ra \un} \abs{v_{ij}(\tau,\cdot)}_{m,M_0}=0\qq\A\,m\in \N,
\end{equation}
where $\abs{\cdot}_{m,M_0}$ denotes the norm in $C^m(M_0)$.
\et

\section{Spacelike slices in the exterior region}\las{2}

For our purposes, it is best to express the metric in the exterior region of an odd-dimensional Kerr-AdS black hole $N=N^{n+1}$, $n=2m$, in generalized Boyer-Lindquist coordinates. Recall that all rotational parameters are assumed to be equal, $a_i=a\not=0$, $1\le i\le m$. The metric then has the following form, \cf \cite[equ. (8)]{gibbons:bl}; we refer to the arXiv version because the published version contains a misprint:
\begin{equation}\lae{8.2.108}
\begin{aligned}
d\bar s^2&= -\frac{1+l^2r^2}{1-a^2l^2}dt^2+\frac{U r^2 dr^2}{(1+l^2r^2)(r^2+a^2) U -2 m_0 r^2}\\
&\q +\frac{r^2+a^2}{1-a^2l^2}\sum_{i=1}^m(d\m_i^2+\m_i^2d\f_i^2)\\
&\q+\frac{2m_0}U\big(\frac1{1-a^2l^2}dt-\frac a{1-a^2l^2}\sum_{i=1}^m\m_i^2d\f^i\big)^2,
\end{aligned}
\end{equation}
where
\begin{equation}
U=(r^2+a^2)^{m-1},
\end{equation}
\begin{equation}
a>0,
\end{equation}
\begin{equation}
0<r_2<r<\un,
\end{equation}
\begin{equation}
\F(r)={(1+l^2r^2)(r^2+a^2) U -2 m_0 r^2}
\end{equation}
has the property that
\begin{equation}
0=\F(r_2)<\F(r)\qq\A\,r\in  (r_2, \un);
\end{equation}
$r_2$ is the radius of the event horizon, $m_0$ is the mass of the black hole, and $l^2$ is a multiple of the cosmological constant $\Lam$
\begin{equation}\lae{2.7.1}
l^2=-\frac 1{m(2m-1)}\Lam,
\end{equation}
where we assume $\Lam<0$.
For nonextremal black holes, there are two horizons with radii $r_1$ and $r_2$ such that $0<r_1<r_2$.
Moreover, the "coordinates" $\mu_i$ satisfy
\begin{equation}\lae{2.7}
\sum_{i=1}^m\mu_i^2=1.
\end{equation}
The given coordinates $(t,r,\mu_i,\f^i)$, $1\le i\le m$, represent $n+1=2m+1$ coordinates because of \re{2.7}. We use the symbol $t$ for the time variable rather than the more common symbol $\tau$---$t$ usually denotes the time variable in the Kerr-Schild coordinate system, whereas $\tau$ plays that role in the Boyer-Lindquist system---because $\tau$ will have a different meaning in this paper.

This misprint also appears in \cite[equ. (8)]{gibbons:kerr-ads}, which led us to reproduce it in our book \cite[equ. (8.4.1)]{cg:qgravity-book2}. Fortunately, it had no consequences, because we were interested in spacelike slices in the interior region of the Kerr-AdS black hole, where we used the correctly printed Kerr-Schild coordinates.

The term
\begin{equation}
\sum_{i=1}^m d\m_i^2
\end{equation}
in equation \re{8.2.108} represents the standard metric on $\Ss[m-1]$, and 
\begin{equation}\lae{2.10}
\sum_{i=1}^m (d\mu_i^2+\mu_i^2d\f_i^2),
\end{equation}
is the standard metric on $\Ss[2m-1]$, where the $\f^i$ are azimuthal coordinates whose values must be identified modulo $2\pi$.
 
Indeed, let $\mu^i(\theta^a)$, $1\le a\le m-1$, be a local embedding of the standard sphere $\Ss[m-1]$ into $\R[m]$ given by
\begin{equation}
\mu=(\mu^i): \Om \su \R[m-1]\longrightarrow \R[m]
\end{equation}
and satisfying
\begin{equation}\lae{2.12.1}
\abs\mu^2=1.
\end{equation}
Then
\begin{equation}\lae{2.14.1}
\s_{ab}=\de_{ij}\mu^i_a\mu^j_b
\end{equation}
is the induced metric and
\begin{equation}\lae{2.14.2} 
\s_{ab}d\theta^a d\theta^b=\sum_{i=1}^m d\mu_i^2.
\end{equation}
The coordinates $(\theta^a,\f^i)$ are therefore polyspherical coordinates on $\Ss[2m-1]$ associated with a special tree structure, \cf \cite[p.~225]{vilenkin_1993}.

The metric in \re{8.2.108} can be expressed as 
\begin{equation}\lae{8.2.15.1}
\begin{aligned}
d\bar s^2&= -\frac{1+l^2r^2}{1-a^2l^2}dt^2+\frac{U r^2 dr^2}{(1+l^2r^2)(r^2+a^2) U -2 m_0 r^2}\\
&\q +\frac{r^2+a^2}{1-a^2l^2} \s_{ab}d\theta^a d\theta^b+\frac{r^2+a^2}{1-a^2l^2}\sum_{i=1}^m\m_i^2d\f_i^2\\
&\q+\frac{2m_0}U\big(\frac1{1-a^2l^2}dt-\frac a{1-a^2l^2}\sum_{i=1}^m\m_i^2d\f^i\big)^2.
\end{aligned}
\end{equation}
Using the abbreviations
\begin{equation}\lae{2.16.1}
A=\frac{2m_0}U\frac1{1-a^2l^2}\big(\frac1{1-a^2l^2}-(1+l^2r^2)\big),
\end{equation}
\begin{equation}\lae{2.17.1}
B=-\frac{2m_0}U\frac{a}{(1-a^2l^2)^2},
\end{equation}
\begin{equation}\lae{2.18.1}
C=\frac{r^2+a^2}{1-a^2l^2},
\end{equation}
\begin{equation}\lae{2.19.1}
b^2=\frac{2m_0}U\frac{a^2}{1-a^2l^2}\frac1{r^2+a^2}
\end{equation}
and
\begin{equation}\lae{2.20.1}
h^{-1}=\frac{U r^2 }{(1+l^2r^2)(r^2+a^2) U -2 m_0 r^2},\qq\A\, r>r_2,
\end{equation}
the metric above is equal to
\begin{equation}\lae{2.21}
\begin{aligned}
d\bar s^2&=A dt^2+2B\mu_i^2d\f^i dt+C\big(\mu_i^2\de_{ij}+b^2\mu_i^2\mu_j^2\big)d\f^i d\f^j\\
&\q +C \s_{ab}d\theta^a d\theta^b+d\tau^2
\end{aligned}
\end{equation}
in the exterior of the black hole, where the new coordinate $\tau$, $0\le\tau<\un$, is defined by
\begin{equation}\lae{2.22}
\tau=\int_{r_2}^r h(s)^{-\frac12}\,ds
\end{equation}
so that
\begin{equation}\lae{2.23}
\pde r\tau=h^\frac12.
\end{equation}
We denote differentiation with respect to $\tau$ by a prime. Then
\begin{equation}\lae{2.24.1}
r'=h^\frac12>0,
\end{equation}  
\begin{equation}\lae{2.25.1}
r''=\frac12 \pde hr,
\end{equation}
\begin{equation}\lae{2.26.1}
\ga_1\abs\Lam\le r^{-1}\frac12\pde hr=l^2+O(r^{-2})\qq\A\, r\ge r_3>>r_2,
\end{equation}
where $\ga_1$ is a uniform constant independent of $\Lam$,
\begin{equation}\lae{2.27.1}
r^{-2}h=l^2+O(r^{-2})\qq\A\, r\ge r_3>>r_2
\end{equation}
and 
\begin{equation}
\pde{\log r}\tau=r^{-1}h^\frac12\ge \ga_0 \abs{\Lam}\qq\A\, r\ge r_3>>r_2,
\end{equation}
where $\ga_0$ is independent of $\Lam$. These estimates follow immediately from the definition \re{2.20.1}. We further deduce
\begin{equation}\lae{2.27.2}
r(\tau)^{-1}\le c_2 e^{-c_1\tau}\qq\A\,\tau\ge \tau_3,
\end{equation}
where
\begin{equation}
r(\tau_3)=r_3.
\end{equation}
Here, $c_1$ and $c_2$ are uniform positive constants.

We shall need the inverse of the metric in \re{2.21} for later use. Given its block structure, it is evident that the only nontrivial block is 
\begin{equation}\lae{2.29.1}
d s^2=A dt^2+2B\mu_i^2d\f^i dt+C\big(\mu_i^2\de_{ij}+b^2\mu_i^2\mu_j^2\big)d\f^i d\f^j.
\end{equation}
Let us first invert the metric
\begin{equation}\lae{2.30.1}
g_{ij}=\mu_i^2\de_{ij}+b^2\mu_i^2\mu_j^2,\qq 1\le i,j\le m,
\end{equation}
by applying the Sherman-Morrison formula:
\bl
Let $A\in L(\R[m],\R[m])$ be invertible, and let $u,v\in \R[m]$. Then the linear map
\begin{equation}\lae{2.31.1} 
A+uv^T,
\end{equation}
is also invertible if and only if
\begin{equation}\lae{2.32.1}
1+v^TA^{-1}u\not=0,
\end{equation}
with inverse 
\begin{equation}\lae{2.33.1}
A^{-1}-\frac{A^{-1}uv^TA^{-1}}{1+v^TA^{-1}u},
\end{equation}
where the superscript $T$ denotes transposition, the product in \re{2.31.1} is a dyad, and the product in \re{2.32.1} is the same as the scalar product
\begin{equation}
\spd{v}{A^{-1}u}.
\end{equation}
\el
\bp
See, for example, \cite[Section 2]{sherman-morrison}.
\ep
Using the Sherman-Morrison formula, we infer
\bl\lal{2.2}
The inverse $(g^{ij})$ of the metric $(g_{ij})$ is
\begin{equation}
g^{ij}=\mu_i^{-2}\de^{ij}-\frac{b^2}{1+b^2}e^{ij},
\end{equation}
where
\begin{equation}
e^{ij}=1,\qq\A\, 1\le i,j\le m.
\end{equation}
\el
\bp
The matrix $A$ is diagonal,
\begin{equation}
A=\diag (\mu_1^2,\ldots,\mu_m^2),
\end{equation}
where $\mu_i^2>0$ and the vectors $u,v$ are defined by
\begin{equation}\lae{2.38.1}
u=v=b x,
\end{equation}
where $x=(\mu_i^2)$. Hence, condition \re{2.32.1} becomes
\begin{equation}\lae{2.39.1}
1+v^TA^{-1}u=1+b^2\not=0,
\end{equation}
in view of \re{2.12.1}, and the quotient in \re{2.33.1} is equal to
\begin{equation}
\frac{b^2}{1+b^2}.
\end{equation}
\ep
From the determinant lemma for rank-one updated matrices, we then conclude
\bl\lal{2.3}
Let $(g_{ij})$ be the metric in \re{2.30.1}. Then
\begin{equation}
\det g_{ij}=\prod_{i=1}^m\mu_i^2 (1+b^2).
\end{equation}
\el
\bp
The proof follows immediately from \cite[Lemma 1.1, p.~1224]{determinant} and \re{2.39.1}. 
\ep

We are now ready to derive the inverse of the metric in \re{2.29.1}. Let $G$ be the matrix
\begin{equation}
G=C(g_{ij}),
\end{equation}
where the coefficients $g_{ij}$ are defined in \re{2.30.1}, and let $x\in \R[m]$ be the column vector $x=(\mu_i^2)$. Then the metric $(\bar g_{\al\bet})$, $0\le\al,\bet\le m$, can be written in block form as
\begin{equation}
(\bar g_{\al\bet})=
\begin{pmatrix}
A&Bx^T\\[\cma]
Bx&G
\end{pmatrix},
\end{equation}
where $\bar g_{00}=A$, $\bar g_{0j}=B x_j$, $\bar g_{i0}=B x_i$, and $\bar g_{ij}=Cg_{ij}$. By \rl{2.2}, $G$ is invertible, so we can define the Schur complement $S$ of $G$ in $(\bar g_{\al\bet})$ by
\begin{equation}
S=A-B^2x^TG^{-1}x,
\end{equation}
\cf \cite[Section 1]{schur-complement}. In our case
\begin{equation}
x^TG^{-1}x=\frac1{C(1+b^2)},
\end{equation}
i.e.,
\begin{equation}\lae{2.46.1}
S=\frac{AC(1+b^2)-B^2}{C(1+b^2)}\equiv \frac{\varTheta}{C(1+b^2)}. 
\end{equation}
Here $S$ is a real number, but in general it will be a matrix.

We can now apply the block matrix inversion formula:
\bl\lal{2.4}
If $S$ is invertible, or equivalently if $\varTheta\not=0$, then $(\bar g_{\al\bet})$ is invertible with inverse 
\begin{equation}
(\bar g^{\al\bet})=
\begin{pmatrix}
S^{-1} & -S^{-1}(Bx)^TG^{-1}\\
-G^{-1}(Bx)S^{-1} & G^{-1}+G^{-1}(Bx)S^{-1}(Bx)^TG^{-1}
\end{pmatrix}.
\end{equation}
\el
\bp
See, for example, \cite[Section 1]{schur-complement}.
\ep
\bc\lac{2.5}
The components $\bar g^{00}$ and $\bar g^{i0}=\bar g^{0i}$ are 
\begin{equation}
\bar g^{00}=\frac{C(1+b^2)}{\varTheta}
\end{equation}
and
\begin{equation}
\bar g^{i0}=\bar g^{0i}=-\frac B\varTheta, \qq\A\, 1\le i\le m.
\end{equation}
\ec
\bp
This follows by a straightforward computation.
\ep
\br\lar{2.6}
From the definition of $\varTheta$ in \re{2.46.1}, we deduce that the zeros of $\varTheta$ are also the zeros of a polynomial in $r$. Hence, $\varTheta$ has at most finitely many zeros. Moreover, $\varTheta$ is always strictly negative if $A<0$, which is equivalent to
\begin{equation}
r^2>\frac{a^2}{1-a^2l^2},
\end{equation}
in view of \re{2.16.1}.
\er
After these preparations, we can prove the following theorem.
\bt\lat{2.7}
Let $\socc$ be a spacelike slice in the exterior region of the Kerr-AdS black hole, given for fixed $t$ by
\begin{equation}\lae{2.50.1}
\socc=\{x^0=t=\const\}.
\end{equation}
Then the vector field
\begin{equation}\lae{2.51.1}
\nu^\al=\abs{\bar g^{00}}^{-\frac12}\bar g^{0,\al}
\end{equation}
is a unit normal to $\socc$, where $(x^\al)$ denotes the $2m+1$ coordinates
\begin{equation}\lae{2.52.1}
(x^\al)=(t,\f^i,\theta^a,r).
\end{equation}
Here $x^0=t$, the indices corresponding to $\f^i$ range from $1$ to $m$, the $\theta^a$ are coordinates on the unit sphere $\Ss[m-1]$, and $r_2<r<\un$ is the radial coordinate, which may be replaced by $\tau$, \cf \re{2.22}. The only nonzero components of $\nu^\al$ are those with indices
\begin{equation}\lae{2.53.1} 
\al\in \{0,1,\ldots,m\};
\end{equation}
all other components vanish identically. The second fundamental form of $\socc$ vanishes identically, and the scalar curvature $R$ of $\socc$ is constant
\begin{equation}\lae{2.55.1}
R=2\Lam.
\end{equation}
\et
\bp
We first prove \re{2.51.1}. The coordinate $x^0$ can also be considered to be a time function 
\begin{equation}
f(x)=x^0
\end{equation}
 in the exterior region $N_{\tup{ext}}$. Hence,
 \begin{equation}
\socc=f^{-1}(t)
\end{equation}
and 
\begin{equation}
\grad f_{|_\socc}=(1,0,\ldots,0)
\end{equation}
is a covariant normal vector of $\socc$ with norm
\begin{equation}
\norm{\grad f_{|_\socc}}^2=\bar g^{\al\bet}f_\al f_\bet=\bar g^{00}<0,
\end{equation}
since $\socc$ is a spacelike hypersurface. The contravariant normal is
\begin{equation}
f^{\al}=\bar g^{\al\bet}f_\bet=\bar g^{0\al}
\end{equation}
and
\begin{equation}
\nu^\al=\abs{\bar g^{00}}^{-\frac12}\bar g^{0\al}
\end{equation}
is a unit normal. The statement \re{2.53.1} then follows from \re{2.21} and \rc{2.5}.

Before considering the second fundamental form, we use the symbols $a',b',\dots$ for the nonzero indices in the coordinate system $(x^\al)$; thus, the $x^{a'}$ are the spatial coordinates on the Cauchy hypersurface $\socc$. The second fundamental form $h_{a'b'}$ is then defined by the Gaussian formula
\begin{equation}\lae{2.63.1}
x^\al_{a'b'}=x^\al_{,a'b'}-\C^{c'}_{a'b'}x^\al_{c'}+\bar\C^\al_{\bet\ga}x^\bet_{a'}x^\ga_{b'} =h_{a'b'}\nu^\al,
\end{equation}
where a comma indicates partial differentiation. The tensor on the left-hand side represents the full second covariant derivative of the embedding
\begin{equation}\lae{2.64.1}
x^\al(t,x^{a'}): (x^{a'})\in U\ra N_{\tup{ext}},
\end{equation}
of a coordinate chart $(U,(x^{a'}))$ on $\socc$ into $N_{\tup{ext}}$. For details, see, e.g., \cite[Theorem 1.1.2, p.~2]{cg:cp}.

Evaluating \re{2.63.1} for $\al=0$, we infer 
\begin{equation}\lae{2.65.1}
\nu^0 h_{a'b'}=\bar\C^0_{\bet\ga}x^\bet_{a'}x^\ga_{b'} =\bar\C^0_{a'b'},
\end{equation}
in view of definition \re{2.64.1}. Before computing the Christoffel symbols, note that the coefficients of the metric in \re{2.21} do not depend on the variables $x^0$ and $\f^i$ and that
\begin{equation}\lae{2.66.1}
\bar\C^0_{a'b'}=\bar g^{00}[a'b',0]+\bar g^{0k}[a'b',k], 
\end{equation}
since the other components of $\bar g^{0\al}$ vanish. Here $k$ refers to the variable $\f^k$ and the brackets on the right-hand side are the Christoffel symbols of the first kind. It follows directly that the only components that may be nonzero are $h_{ia}=h_{ai}$ and $h_{\tau i}=h_{i\tau}$, and we further infer
\begin{equation}\lae{2.67.1}
\nu^0 h_{ia}=\bar g^{00}\frac12 \bar g_{0i,a}+\bar g^{0k} \frac 12 \bar g_{ik,a}
\end{equation}
and
\begin{equation}\lae{2.68.1}
\nu^0 h_{i\tau}=\bar g^{00}\frac12 \bar g_{0i,\tau}+ \bar g^{0k} \frac 12 \bar g_{ik,\tau}.
\end{equation} 
Applying \re{2.21} and \rc{2.5} yields
\begin{equation}
\nu^0 h_{ia}=\big (\frac{C(1+b^2)}\varTheta B -\frac {BC}\varTheta (1+b^2)\big ) \mu_i \mu_{i,a}=0
\end{equation}
and
\begin{equation}
\nu^0 h_{i\tau}=\big (\frac{C(1+b^2)}\varTheta B -\frac {BC}\varTheta (1+b^2)\big )_\tau\,\mu_i^2=0.
\end{equation}
Hence, the second fundamental form of $\socc$ vanishes identically except possibly at finitely many points. Because $\socc$ is a smooth hypersurface, it also vanishes at those exceptional points.

It remains to prove \re{2.55.1}. Since the black hole satisfies the Einstein equations
\begin{equation}
G_{\al\bet}+\Lam \tilde g_{\al\bet}=0
\end{equation}
and $h_{a'b'}=0$, we deduce from the Gauss equation
\begin{equation}
R=-\{H^2-\abs A^2\}+2G_{\al\bet}\nu^\al\nu^\bet=2G_{\al\bet}\nu^\al\nu^\bet=2\Lam,
\end{equation}
\cf \cite[equ. (1.1.43), p.\,5]{cg:cp}, completing the proof of the theorem.
\ep 

Let us now analyze the induced metric of $\socc$. From \re{2.21} we immediately deduce
\begin{equation}\lae{2.72.1}
\begin{aligned}
ds^2&=d\tau^2+ C\s_{ab} d\theta^a d\theta^b + C(\mu_i^2\de_{ij}+b^2\mu_i^2\mu_j^2)d\f^id\f^j\\
&=d\tau^2+C(\s_{ab} d\theta^a d\theta^b + g_{ij}d\f^id\f^j)\\
&\equiv d\tau^2+\rho^2g_{\al\bet}d\xi^\al d\xi^\bet,
\end{aligned}
\end{equation}
where $(\xi^\al)=(\theta^a,\f^i)$ are coordinates on the unit sphere $\Ss[2m-1]$ with corresponding metric
\begin{equation}\lae{2.73.1}
\tilde \s_{\al\bet}d\xi^\al d\xi^\bet= \s_{ab} d\theta^a d\theta^b+\mu_i^2 (d\f^i)^2,
\end{equation}
\cf \re{2.10}, \re{2.14.1}, and \re{2.30.1}. Here,
\begin{equation}\lae{2.74.1}
\rho^2=C=\frac{r^2+a^2}{1-a^2l^2},
\end{equation}
as defined in \re{2.18.1}. 

Applying \rl{2.2}, we derive
\begin{equation}\lae{2.75.1}
\begin{aligned}
g&=\det g_{\al\bet}=\det \s_{ab}\det g_{ij}=\det \s_{ab} \prod _{i=1}^m \mu_i^2 (1+b^2)\\
&=(1+b^2)\det \tilde \s_{ab}=(1+b^2) \tilde\s.
\end{aligned}
\end{equation}
Let $\D_g$ be the Laplacian of the metric above, and let $\psi\in C^2(\Ss[2m-1])$. Then
\begin{equation}
\begin{aligned}
-\D_g\psi&=-\frac1{\sqrt {\tilde\s}}\pd{}\xi\al\big(\sqrt{\tilde\s} g^{\al\bet} \psi_{\xi^\bet} \big)\\
&=-\frac1{\sqrt {\tilde\s}}\pd{}\theta a\big(\sqrt {\tilde\s} \s^{ab}\psi_{\theta^b}\big)-\frac1{\sqrt {\tilde\s}}\pd{}\f i\big(\sqrt {\tilde\s} g^{ij}\psi_{\f^j}\big)\\
&=-\frac1{\sqrt {\tilde\s}}\pd{}\xi\al\big(\sqrt{\tilde\s} \tilde\s^{\al\bet} \psi_{\xi^\bet} \big)+\frac{b^2}{1+b^2}e^{ij}\psi_{ij}\\
&=-\D_{\tilde\s}\psi+\frac{b^2}{1+b^2}e^{ij}\psi_{ij},
\end{aligned}
\end{equation}
in view of \rl{2.2}, where $\psi_{ij}$ are the mixed partial derivatives of $\psi$ with respect to $\f^i$ and $\f^j$.

We seek to determine the eigenfunctions $\psi$ of $-\D_g$. It is well known that the eigenfunctions $\psi$ of $-\D_{\tilde \s}$, i.e., the eigenfunctions on $\Ss[2m-1]$, can be written as products of functions $F_\al$, each depending on only one coordinate $\xi^\al$
\begin{equation}
\psi(\xi)=\prod_{\al=1}^{2m-1}F_\al(\xi^\al),
\end{equation}
provided that these coordinates have a tree structure, \cf \cite[equ. (2), p.~228]{vilenkin_1993}. We assume that the coordinates $(\xi^\al)=(\theta^a,\f^i)$ have such a structure. The coordinates $\f^i$ are of particular interest because they correspond to nodes with open-ended edges and range over $[0,2\pi)$; see figure~(4) on p.~225 and the first figure at the bottom of the same page in \cite{vilenkin_1993}. The corresponding function $F_i(\f^i)$ is
\begin{equation}
F_i(\f^i)=\frac1{\sqrt{2\pi}}e^{i\,m_i\f^i},
\end{equation}
where $m_i\in\Z$. Note that $m_i$ depends on the eigenvalue $\lam$ of $\psi$ and vice versa. Hence, if the eigenvalues $\lam_k$ of $-\D_{\tilde \s}$ are ordered 
\begin{equation}
0=\lam_0<\lam_1\le \lam_2\le \cdots,
\end{equation}
including their multiplicities, the corresponding values $m_i$ are denoted by $m_i(\lam_k)$ or simply by $m_{ik}$. 

Thus, we have proved the following lemma.
\bl\lal{2.8} Let $\psi_k$ be eigenfunctions on $\Ss[2m-1]$ with corresponding eigenvalues $\lam_k$. Then the $\psi_k$ are also eigenfunctions of $-\D_g$ with eigenvalues 
\begin{equation}
\lam_k-\frac{b^2}{1+b^2}\sum_{i=1}^m\sum_{j=1}^m m_{ik}m_{jk},
\end{equation}
i.e., the eigenvalues depend on $b^2$, or equivalently, on $\rho^2$.
\el

Now let $\tilde g$ be the metric in \re{2.72.1}, let $\D$ be its Laplacian, set $n=2m=\dim\socc$, and let $v=v(\tau,\xi)$ be a function in $C^2(\socc)$. Then we deduce
\begin{equation}\lae{2.15}
-\D v=-\frac1{\rho^{n-1}}\frac{\pa}{\pa\tau}\big(\rho^{n-1}v'\big)-\rho^{-2}\D_g v,
\end{equation}
where the prime indicates partial differentiation with respect to $\tau$.

We solve the eigenvalue equation 
\begin{equation}\lae{2.16}
-\D v=\mu v
\end{equation}
by separation of variables. Let $\psi$ be an eigenfunction of $-\D_g$ satisfying
\begin{equation}\lae{2.17}
- \D _g\psi=\big(\lam -\frac{b^2}{1+b^2}\sum_{i=1}^m\sum_{j=1}^m m_i(\lam)m_j(\lam)\big)\psi.
\end{equation}
We define $v$ by
\begin{equation}\lae{2.18}
v=\vt(\tau)\psi(\xi).
\end{equation}
Then $v$ is an eigenfunction if and only if $\vt$ satisfies 
\begin{equation}\lae{2.20}
-\frac1{\rho^{n-1}}\frac{\pa}{\pa\tau}\big(\rho^{n-1}\vt'\big)+\rho^{-2}\big(\lam -\frac{b^2}{1+b^2}\sum_{i=1}^m\sum_{j=1}^m m_i(\lam)m_j(\lam)\big)\vt=\mu \vt.
\end{equation}
For ease of reference, define
\begin{equation}\lae{2.88.1}
\tilde\lam=\big(\lam -\frac{b^2}{1+b^2}\sum_{i=1}^m\sum_{j=1}^m m_i(\lam)m_j(\lam)\big),
\end{equation}
and
\begin{equation}\lae{2.87.1}
A_{n-1}\vt=-\frac1{\rho^{n-1}}\frac{\pa}{\pa\tau}\big(\rho^{n-1}\vt'\big)+\rho^{-2}\tilde\lam\vt=\mu\vt.
\end{equation}
The operator $A_{n-1}$ may be regarded as a densely defined symmetric operator on a suitable Hilbert space.
\bd 
Let $I=(0,\un)$, and for real-valued functions $\vt_1,\vt_2\in C^\un_c(I)$ define the scalar product
\begin{equation}
\spd{\vt_1}{\vt_2}_{n-1}=\int_I\vt_1\vt_2 \rho^{n-1}d\tau,
\end{equation}
where $\rho=\rho(r(\tau))=\rho(\tau)$, \cf \re{2.74.1}, and $r(\tau)$ is the function implicitly defined in \re{2.22}. We denote the completion of this scalar-product space by $L^2(I,n-1)$. To display the measure explicitly, we sometimes write $L^2(I,\rho^{n-1}d\tau)$. 

The operator $A_{n-1}$ defined in \re{2.87.1} is a densely defined symmetric operator with corresponding bilinear form
\begin{equation}
\spd{\vt_1}{\vt_2}_2=\spd{A_{n-1}\vt_1}{\vt_2}_{n-1}\qq\A\, \vt_1,\vt_2\in C^\un_c(I).
\end{equation}
The right-hand side of this equation can be written as an integral 
\begin{equation}
\int_I\{\vt_1'\vt_2' \rho^{n-1}+\tilde\lam \rho^{n-3}\vt_1\vt_2\}d\tau=\spd{\vt_1'}{\vt_2'}_{n-1}+\tilde\lam\spd{\rho^{-2}\vt_1}{\vt_2}_{n-1}.
\end{equation}
\ed
Starting from this definition, which is analogous to \cite[Definition 1, p.~7]{cg:qbh25}, we can apply the subsequent results and arguments of \cite{cg:qbh25} to the present setting by replacing $r,\psi,\lam$ with $\rho,\vt,\tilde\lam$, respectively, and making a few other minor adjustments. To keep the present paper readable and reasonably self-contained, we repeat and reformulate the necessary definitions and results without detailed proofs, referring instead to our previous paper.
\bl
The operator $A_{n-1}$ is bounded below
\begin{equation}
A_{n-1}\ge -c_0
\end{equation}
for sufficiently large $c_0=c_0(\rho_2,\tilde\lam)$.
\el 
\bp
This is immediate because 
\begin{equation}
0<\rho_2=\rho(0)\le \rho(\tau)\qq\A\,\tau\in I.
\end{equation}
\ep
Hence, $A_{n-1}$ admits a self-adjoint extension, namely the Friedrichs extension, \cf \cite[Chapter V.4, p.~110]{maurin:book}.

\br
The preceding definitions and results extend immediately to complex-valued functions when the scalar products are defined appropriately.
\er
Next, we define a unitarily equivalent operator whose eigenfunctions or eigendistributions are easier to analyze. First, we write the function $\vt$ in \re{2.87.1} as a product
\begin{equation}\lae{2.26}
\vt=\rho^{-\frac{n-1}2} u,
\end{equation}
where $u=u(\tau)$ is a real-valued function of class $C^2$. Differentiating and simplifying the resulting equation yields
\begin{equation}\lae{2.96.1}
\begin{aligned}
&-u''+\rho^{-2} \big\{\tilde\lam -\frac{n-1}2(1+\frac{n-1}2)\abs{\rho'}^2+\frac{n-1}2\rho \rho''\big\}u=\mu u.
\end{aligned}
\end{equation}
From the definition of $h$ in \re{2.20.1}, equations \re{2.23}--\re{2.27.1}, and the identity
\begin{equation}\lae{2.97.1}
\rho\rho'=r r' (1-a^2l^2)^{-1},
\end{equation}
we deduce
\begin{equation}\lae{2.98.1}
\begin{aligned}
\rho^{-2}\abs{\rho'}^{2}&=\frac{r^4}{\rho^4} r^{-2}\abs{r'}^2(1-a^2l^2)^{-2}\\
&=\frac1{(1+a^2r^{-2})^2}r^{-2}\abs{r'}^2\\
&= l^2+O(r^{-2}).
\end{aligned}
\end{equation}
Differentiating \re{2.97.1} and applying \re{2.98.1}, \re{2.25.1}, and \fre{2.26.1}, we also obtain
\begin{equation}\lae{2.99.1}
\begin{aligned}
\rho^{-1}\rho''&=\big(\{\frac{r^2}{r^2+a^2}r^{-2}\abs{r'}^2-\rho^{-2}\abs{\rho'}^2\}+\frac{r^2}{r^2+a^2}r^{-1}r''\big)\\
&=l^2+O(r^{-2}).
\end{aligned}
\end{equation}
Using these results, equation \re{2.96.1} can be expressed in the form 
\begin{equation}\lae{2.100.1}
\begin{aligned}
\mu u&=-u''+q u-\frac{(n-1)^2}4 l^2 u\\
&=-u''+q u -\frac{n-1}{2n}\abs\Lam u,
\end{aligned}
\end{equation}
in view of \re{2.7.1}, where $q=q(r(\tau))$ satisfies the estimate
\begin{equation}\lae{2.101.1}
\abs q\le c_0r^{-2}\le c e^{-\ga \tau}\qq\A\, \tau\ge0,
\end{equation}
\cf \re{2.98.1}, \re{2.99.1}, and \re{2.27.2}. The positive constants are uniform and depend on $\tilde\lam,n,r_2,m_0$, but not on $\abs\Lam$. In view of \re{2.88.1} $\tilde \lam$ is bounded independently of $b^2$.

The right-hand side of equation \re{2.100.1} defines the linear differential operator $A_0$
\begin{equation}\lae{2.102.1}
A_0u=-u''+q u -\frac{n-1}{2n}\abs\Lam u,
\end{equation}
which is unitarily (or, in the real case, orthogonally) equivalent to $A_{n-1}$ defined in \re{2.87.1}.
\bl\lal{2.5}
Let $L^2(I)=L^2(I,d\tau)$ be the usual Hilbert space of square-integrable real-valued functions $u$ on $I$, with the standard scalar product
\begin{equation}
\spd{u_1}{u_2}=\int_I u_1 u_2d\tau.
\end{equation}
Then the map
\begin{equation}
\begin{aligned}
\F:L^2(I,n-1)&\ra L^2(I)\\
\vt&\ra \rho^{\frac{(n-1)}2}\vt
\end{aligned}
\end{equation}
is orthogonal and the operators $A_{n-1}$ and $A_0$, the domains of which are the test functions $C^\un_c(I)$, are orthogonally equivalent
\begin{equation}
A_{n-1}=\F^{-1}\circ A_0\circ \F.
\end{equation}
\el
\bp
This follows immediately from the arguments after the ansatz \re{2.26}.
\ep

$A_0$ is densely defined, symmetric, and bounded below; hence, it has a self-adjoint Friedrichs extension $\tilde A_0$. The operator $\tilde A_0$ is defined on the Sobolev space $H^{1,2}_0(I)\ii H^{2,2}(I)$. The usual procedure is to extend $A_0$ to this Sobolev space by defining distributional derivatives and proving that $C^\un_c(I)$ is dense with respect to the corresponding norm. Therefore, we shall omit the tilde and regard $A_0$ as defined on $H^{1,2}_0(I)\ii H^{2,2}(I)$. We shall later prove that the eigenvalue equation
\begin{equation}\lae{2.35}
A_0u=\mu u
\end{equation}
has no square-integrable solutions and that $A_0$ has purely continuous spectrum, so that only eigendistributions can be found. For now, we treat the eigenvalue equation simply as an ODE that, for any $\mu\in \R[]$, has a unique solution $u\in C^2([0,\un))$ with initial values
\begin{equation}\lae{2.36}
u(0)=0\q\wed\q u'(0)=1.
\end{equation}
To analyze the solutions of the ODE, we rewrite the eigenvalue equation \re{2.35} as
\begin{equation}\lae{2.39}
-u''+q u=\big(\mu +\frac{n-1}{2n}\abs\Lam\big) u\equiv k^2 u,
\end{equation}
assuming
\begin{equation}
k>0,
\end{equation}
or equivalently,
\begin{equation}\lae{2.41}
\mu>-\frac{n-1}{2n}\abs\Lam.
\end{equation}
It is well known that the eigenvalue equation \re{2.39} has no square-integrable solutions if $k>0$, \cf \cite[Theorem 4.1, p.~14]{simon-spectrum}. In fact, the hypotheses for this result can be weakened, because the condition
\begin{equation}
\limsup _{\tau\ra\un}\tau\abs q=0
\end{equation}
would suffice. In \cite[Theorem 2, p.~9]{cg:qbh25}, we proved the following theorem:
\bt\lat{2.13}
Let $u\in C^2([0,\un))$ be a nontrivial solution of the equation
\begin{equation}\lae{2.48}
-u''+q u=k^2 u
\end{equation}
with $k>0$, where $q$ satisfies estimate \re{2.101.1}. Then
\begin{equation}\lae{2.49}
\abs u^2+k^{-2}\abs {u'}^2\le c_0\qq\A\, \tau\ge 0,
\end{equation}
where $c_0$ depends on $k,c,\ga$ and the initial values of $u$ at $\tau=0$. Moreover, $u$ is not square-integrable.
\et
\br
Let $A$ be the operator defined by the left-hand side of equation \re{2.48}, with domain
\begin{equation}
D(A)=C^\un_c(I),
\end{equation}
where $I=(0,\un)$. Assume $q\ge 0$, which in our case would require the eigenvalue $\tilde\lam$ in \re{2.88.1} to be sufficiently large. Then the quadratic form defined by
\begin{equation}
\spd{Au}{u}=\int_0^\un (\abs {u'}^2+q\abs u^2)d\tau,\qq u\in D(A),
\end{equation}
would be nonnegative and, in fact, positive definite, because
\begin{equation}
\spd{Au}u=0
\end{equation}
implies $u\equiv \const$
and hence $u=0$ because $u(0)=0$. It is well known that the self-adjoint extension of $A$, denoted by $\tilde A$, has domain
\begin{equation}
D(\tilde A)=H^{1,2}_0(I)\ii H^{2,2}(I),
\end{equation}
where the function spaces on the right-hand side are the usual Sobolev spaces. Any eigenvector $u$ of $\tilde A$ satisfying
\begin{equation}
\tilde A u=0
\end{equation}
would also belong to $C^2([0,\un))$, satisfy $u(0)=0$, and solve the eigenvalue equation \re{2.48} with $k=0$. Hence, $u$ would vanish identically because
\begin{equation}
\spd {\tilde Au}u=\int_0^\un \tilde Au u \,d\tau=0.
\end{equation}
On the other hand, any square-integrable function $u\in C^2([0,\un))$ satisfying $u(0)=0$ that solves equation \re{2.48} with $k=0$ is an element of $D(\tilde A)$, \cf \cite[Lemma 4.2, p.\,14]{simon-spectrum}. Thus, we conclude $u=0$. 
\er
As a corollary of \rt{2.13}, we obtain
\bc\lac{2.8}
Let $\mu$ satisfy \re{2.41}. The eigenvalue problem in \re{2.35},
\begin{equation}
A_0u=\mu u,
\end{equation}
with initial values $u(0)=0$ and $u'(0)=1$ has no square-integrable solutions. As an ODE, it has a unique solution $u\in C^2([0,\un))$ that satisfies estimate \re{2.49}, where
\begin{equation}\lae{2.66}
k^2=(\mu+\frac{n-1}{2n}\abs\Lam),
\end{equation}
in view of \re{2.39}. Hence, the solution of the ODE is a tempered distribution and may be regarded as an eigendistribution of the unique self-adjoint extension $\tilde A_0$.
\ec

Combining the preceding corollary, \rl{2.5}, and equations \re{2.15}--\re{2.20}, we obtain the following theorem.
\bt\lat{2.9}
The eigenvalue equation \re{2.16}
\begin{equation}
-\D v=\mu v
\end{equation}
can be solved by defining
\begin{equation}
v=\rho^{-\frac{n-1}2} u\psi,
\end{equation}
where $\psi=\psi(\xi)$ is an eigenfunction of equation \re{2.17} and $u$ is a tempered eigendistribution of equation \re{2.35}, provided $\mu$ satisfies \re{2.41}.
\et
\newpage
\nd
To illustrate the behavior of $v$, we include a Mathematica-generated plot in which the eigenfunction solves an equation structurally equivalent to \re{2.16} and an equivalent damping factor is used.
\vskip 1cm
\nd
\begin{figure}[h]
\includegraphics[width=0.956\textwidth]{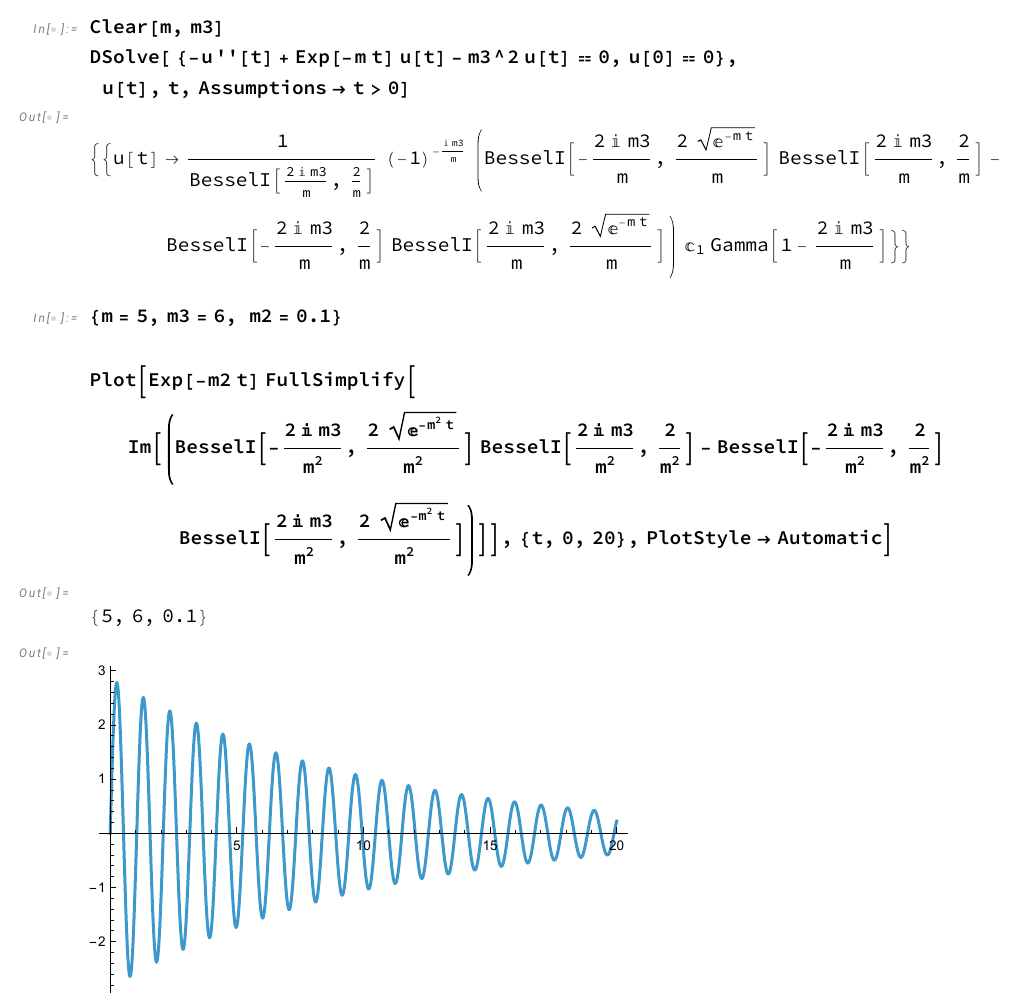}
\end{figure}

\section{Solving the hyperbolic equation}\las{3} 
Recall from the introduction that the final equation associated with the quantized spacetime $N$ is the hyperbolic equation
\begin{equation}\lae{3.1}
H_0{\tilde u}-H_1{\tilde u}=0
\end{equation}
in a quantum spacetime
\begin{equation}\lae{3.2}
Q=(0,\un)\times \so
\end{equation}
where $\so$ is a Cauchy hypersurface of $N$ with induced metric $\tilde g$. The temporal Hamiltonian $H_0$ is a self-adjoint operator with countably many eigenfunctions $w_i$ and corresponding eigenvalues $\lam_i>0$. In our case, the spatial Hamiltonian $H_1$ is also self-adjoint but has continuous spectrum.

To solve the hyperbolic equation, we must solve the eigenvalue equation
\begin{equation}\lae{3.3}
H_1v_i=\lam_i v_i,
\end{equation}
for every $i\in\N$, where
\begin{equation}\lae{3.4}
H_1v=\frac {16(n-1)}n \{-(n-1)\D v-\frac{n-2}2 Rv\}
\end{equation}
and the Laplacian and scalar curvature refer to the induced metric on the Cauchy hypersurface in the exterior region of the black hole
\begin{equation}
\so=\{t=\const\}.
\end{equation}
Recall that the induced metric is independent of $t$ and that the scalar curvature $R$ is constant
\begin{equation}\lae{3.6}
R=2\Lam,
\end{equation}
\cf \frt{2.7}. The solutions $\tilde u$ of \re{3.1} can therefore be expressed as products of the corresponding eigenfunctions or eigendistributions
\begin{equation}
\tilde u_i=w_iv_i\qq\A\,i\in \N.
\end{equation}
The temporal eigenvalues $\lam_i$ all have multiplicity one, whereas the spatial eigenvalues have higher multiplicities, as will become evident shortly. Hence, we label the spatial eigenfunctions in the form
\begin{equation}
v_{ij},\qq 1\le j\le m_i,
\end{equation}
such that
\begin{equation}\lae{3.9}
H_1v_{ij}=\lam_i v_{ij}, \qq\A\,1\le j\le m_i.
\end{equation}
The eigenfunctions $v$ in \frt{2.9} with eigenvalue $\mu$ are also eigenfunctions of $H_1$, because $R$ is constant. Hence, we obtain
\begin{equation}\lae{3.10}
\begin{aligned}
H_1v&= \frac {16(n-1)}n\{(n-1)\mu +(n-2)\abs\Lam\}v\\
&=\frac {16(n-1)}n (n-1)\{\mu +\frac{n-2}{n-1}\abs\Lam\}v,
\end{aligned}
\end{equation}
in view of \re{3.4} and \re{3.6}.

To solve equation \re{3.3} for every $i\in\N$, we must prove that there exists an admissible $\mu\in \R[]$ such that
\begin{equation}
\lam_0\ge \frac {16(n-1)}n (n-1)\{\mu +\frac{n-2}{n-1}\abs\Lam\},
\end{equation}
where $\mu$ is admissible if
\begin{equation}
\mu>-\frac{n-1}{2n}\abs\Lam,
\end{equation}
\cf \fre{2.41}. Thus, we must verify that
\begin{equation}
\lam_0>\frac {16(n-1)}n (n-1)\{-\frac{n-1}{2n} +\frac{n-2}{n-1}\}\abs\Lam.
\end{equation}
On the other hand, we know
\begin{equation}
\lam_i=\bar\lam_i \abs\Lam ^\frac{n-1}n,
\end{equation}
where $\bar\lam_i$ is the eigenvalue corresponding to $\abs\Lam=1$, \cf \cite[Lemma 9.4.8, p.~240]{cg:qgravity-book2}, from which we deduce 
\begin{equation}
\bar\lam_0>\frac {16(n-1)}n (n-1)\{-\frac{n-1}{2n} +\frac{n-2}{n-1}\}\abs\Lam^\frac1n.
\end{equation}
Let $\Lam_0<0$ be defined by
\begin{equation}\lae{3.16}
\bar\lam_0=\frac {16(n-1)}n (n-1)\{-\frac{n-1}{2n} +\frac{n-2}{n-1}\}\abs{\Lam_0}^\frac1n.
\end{equation}
Then we can prove the following theorem.
\bt\lat{3.1}
Let $(m_i)$ be a given sequence of multiplicities with $1\le m_i\in\N$. Then the eigenvalue problems \re{3.9} have solutions of the form
\begin{equation}
v_{ij}=\rho^{-\frac{n-1}2} u_{ij}\psi_{ij},
\end{equation}
where, for each $i\in\N$, the $\psi_{ij}=\psi_{ij}(x)$ are mutually orthogonal eigenfunctions or linearly independent eigendistributions satisfying equation \re{2.17}, and the $u_{ij}$ are linearly independent tempered eigendistributions satisfying equation \re{2.35} and the initial conditions \re{2.36}, provided that the cosmological constant $\Lam$ satisfies
\begin{equation}\lae{3.18} 
\abs{\Lam_0}>\abs\Lam>0.
\end{equation}
\et
\bp
The proof is identical to the proof of \cite[Theorem 4, p.~14]{cg:qbh25}.
\ep

\section{The von Neumann entropy in the exterior region}\las{4}

In \cite[Chapter 9.5, p.~240]{cg:qgravity-book2}, we defined the von Neumann entropy for the interior region of black holes and, more generally, for quantized globally hyperbolic spacetimes with a negative cosmological constant. We now apply these definitions to the exterior region of a Kerr-AdS black hole. For completeness, we repeat the necessary definitions and results without proof; the proofs are given in \cite[Chapter 9.5, p.~240]{cg:qgravity-book2}.

We first define the partition function using the spatial Hamiltonian $H_1$ of the quantized black hole. This Hamiltonian acts on the separable Hilbert space $\mc H$ generated by the eigendistributions
\begin{equation}
v_{ij},\qq \A\, i\in \N,\;1\le j\le m_i,
\end{equation}
which are sufficiently smooth functions satisfying the eigenvalue equations
\begin{equation}
H_1 v_{ij}=\lam_i v_{ij}.
\end{equation}
For the moment, we assume that the multiplicities $m_i$ are uniformly bounded, although we will later drop the requirement
\begin{equation}
m_i\le l,
\end{equation}
where $1\le l\in\N$ is arbitrary but fixed, \cf \rr{4.6}.

To explain how the eigendistributions generate a Hilbert space, we relabel the eigenfunctions and eigenvalues of $H_1$ as $(\tilde v_i,\tlam_i)$ so that
\begin{equation}
H_1 \tilde v_i=\tlam_i \tilde v_i,
\end{equation}
i.e., the eigenvalue multiplicities are now included in the labeling, and the ordering is no longer strict
\begin{equation}
\tlam_0\le \tlam_1\le \tlam_2\le \cdots.
\end{equation}
To define the Hilbert space $\mc H$, we declare the eigendistributions to be mutually orthogonal unit eigenvectors. This declaration defines a scalar product on the complex vector space $\mc H'$ spanned by these eigenvectors; we take the first argument of the scalar product to be antilinear. We then define $\mc H$ as the completion of $\mc H'$.
\bl\lal{4.1}
The linear operator $H_1$ with domain $\mc H'$ is essentially self-adjoint in $\mc H$. Let $\bar H_1$ be its closure. Then the only eigenvectors of $\bar H_1$ are those of $H_1$.
\el
\br
In the following, we shall write $H_1$ instead of $\bar H_1$. 
\er
\bl
For any $\bet>0$, the operator
\begin{equation}
e^{-\bet H_1}
\end{equation}
is of trace class in $\mc H$. Let
\begin{equation}
\msc F\equiv\msc F_+(\mc H)
\end{equation}
be the symmetric Fock space generated by $\mc H$, and let
\begin{equation}
H=d\C(H_1)
\end{equation}
be the canonical extension of $H_1$ to $\msc F$. Then
\begin{equation}
e^{-\bet H}
\end{equation}
is also of trace class in $\msc F$
\begin{equation}\lae{4.12}
\tr(e^{-\bet H})=\prod_{i=0}^\un(1-e^{-\bet \tlam_i})^{-1}<\un.
\end{equation}
\el
We then define the partition function $Z$ by
\begin{equation}
Z=\tr(e^{-\bet H})=\prod_{i=0}^\un (1-e^{-\bet\tlam_i})^{-1},
\end{equation}
and the density operator $\rho$ in $\msc F$ by
\begin{equation}
\rho=Z^{-1}e^{-\bet H},
\end{equation}
so that
\begin{equation}
\tr \rho=1.
\end{equation}

The von Neumann entropy $S$ is then defined by
\begin{equation}\lae{4.16}  
\begin{aligned}
S&=-\tr(\rho\log \rho)\\
&=\log Z+\bet Z^{-1}\tr (He^{-\bet H})\\
&=\log Z-\bet\pde{\log Z}\bet\\
&\equiv \log Z +\bet E,
\end{aligned}
\end{equation}
where $E$ is the average energy
\begin{equation}
E=\tr (H\rho).
\end{equation}
$E$ can be expressed in the form
\begin{equation}\lae{4.18}
E=\sum_{i=0}^\un \frac{\tlam_i}{e^{\bet\tlam_i}-1}.
\end{equation}
We also set the Boltzmann constant to
\begin{equation}
K_B=1.
\end{equation}
The parameter $\bet$ is the inverse of the absolute temperature $T$
\begin{equation}
\bet=T^{-1}.
\end{equation}
\bl\lal{4.4} 
The average energy $E$ and the entropy $S$ increase with the eigenvalue multiplicities $m_i$. For a fixed index $k$, if the corresponding multiplicity $m_k$ tends to infinity, then both $E$ and $S$ tend to infinity
\begin{equation}
\lim_{m_k\ra\un}E=\un\q\wed\q \lim_{m_k\ra\un}S=\un.
\end{equation}
\el
\bp
In view of \re{4.18}, we infer
\begin{equation}
E=\sum_{i=0}^\un m_i \frac{\lam_i}{e^{\bet\lam_i}-1}
\end{equation}
and
\begin{equation}
S\ge \bet E,
\end{equation}
by equation \re{4.16} and the fact that $Z>1$, \cf \re{4.12}; recall that the eigenvalues are strictly positive. The claims of the lemma are now easily deduced.
\ep
We therefore face a dilemma in defining physically reasonable average energy and entropy, because the multiplicities $m_i$ can be increased arbitrarily, as proved in \rt{3.1}. For each $i$, there are countably many mutually orthogonal eigenfunctions $\psi_{ij}$ satisfying equation \fre{2.17}; each can be used to define a solution $u_{ij}$ of the eigenvalue equation \fre{2.35} with eigenvalue $\mu=\lam_i$ without changing that eigenvalue. The eigenvalues corresponding to the $\psi_{ij}$ determine only the coefficients of the lower-order terms of $A_0$, whose continuous spectrum is independent of those eigenvalues. Hence, we cannot develop meaningful quantum statistics for the exterior region by treating it as an isolated system. Instead, we must also take the interior region of the black hole into account.

In \cite[Chapter 8.3, p.~192]{cg:qgravity-book2}, we quantized the interior region of a Kerr-AdS black hole. There, $x^0=r$, with $0<r_1<r<r_2$, is the time function, and the Cauchy hypersurfaces are the slices
\begin{equation}\lae{4.22.1}
S_r=\{x^0=r\}.
\end{equation}
The induced metric is identical to the metric in \fre{2.72.1}, except that $r$, and hence $\rho$, is constant, \cf \cite[equ. (8.2.46), p.~191]{cg:qgravity-book2}. The corresponding Hamiltonian $H_1$ then has the form given in \fre{3.4}, and for any temporal eigenvalue $\lam_i$ we could find spatial eigendistributions $v_{ij}$ satisfying
\begin{equation}\lae{4.23.1}
H_1 v_{ij}=\lam_i v_{ij}\qq\A\, 1\le j\le n(\lam_i),
\end{equation}
where $n(\lam_i)$ is a finite, maximal multiplicity, provided the scalar curvature satisfies
\begin{equation}
R\ge 0,
\end{equation}
so that the smallest eigenvalue $\tilde\mu_0$ of the nontrivial part of the operator $H_1$ is nonpositive. In \cite[discussion following eq.~(9.4.13), p.~229]{cg:qgravity-book2}, we argued that this condition can be achieved by choosing the rotational parameter $a$ sufficiently small, because the scalar curvature converges to that of a round sphere in Euclidean space as $a$ tends to zero. In the cited reference, we treated $R$ simply as a smooth function. However, because the odd-dimensional unit spheres $\Ss[2m-1]$ are Sasakian manifolds, \cf \cite[Section 5, p.~258]{sasaki:normal}, the nonflat part of the induced metric of $S_r$ can be identified with a special deformation of a Sasakian metric; for details, see \frl{5.3.1}. The scalar curvature ${}^*R$ of the deformed metric is then
\begin{equation}\lae{4.25.1}
\begin{aligned}
{}^*R&=R_0-b^2 ((2m-1)-1)\\
&=(2m-1)((2m-1)-1)-b^2((2m-1)-1)\\
&=2(m-1)(2m-1 -b^2),
\end{aligned}
\end{equation}
where $R_0$ is the scalar curvature of $\Ss[2m-1]$ and $b^2$ is defined in \fre{2.19.1} with $r$ fixed. Hence, we conclude that the scalar curvature $R$ of $S_r$,
\begin{equation}\lae{4.26.1}
R=\rho^{-2} \,{}^*R=\rho^{-2} \,2(m-1)(2m-1 -b^2),
\end{equation}
is constant and positive whenever $b^2$ is sufficiently small, which is the case if the rotational parameter $a$ is chosen appropriately. Since we are interested in the limit $r\ra r_2$, it suffices to impose the necessary condition at $r=r_2$. We then have
\begin{equation}
b^2=\frac{a^2r_2^{-2}+a^2l^2}{1-a^2l^2},
\end{equation}
because
\begin{equation}
\frac{2m_0}U \frac1{r_2^2+a^2}=\frac{1+r_2^2l^2}{r_2^2}=r_2^{-2}+l^2.
\end{equation}
Hence, we obtain the following lemma.
\bl\lal{4.5.1}
The scalar curvature $R$ of the radial slices $S_r$ in the interior of the Kerr-AdS black hole has the constant value given in \re{4.26.1}, which is positive provided that
\begin{equation}
b^2=\frac{a^2r_2^{-2}+a^2l^2}{1-a^2l^2}<2m-1,
\end{equation}
and $r_1<r<r_2$ is close to $r_2$. The preceding inequality is satisfied if $a>0$ is sufficiently small, because $r_2$ then tends to the black hole radius of the corresponding Schwarzschild-AdS black hole and hence remains uniformly bounded below by a positive constant as $a\ra 0$.
\el

We can now argue as in \cite[Section 4, p.~16]{cg:qbh25}. The multiplicities $n(\lam_i)$ of the eigenvalues of the corresponding spatial Hamiltonian $H_1$ were defined by maximizing their values. This maximal-multiplicity prescription remains valid as the slices $S_r$ converge to the event horizon, i.e., as $r$ converges to $r_2$. Furthermore, for $\bet>0$, the operator
\begin{equation}
e^{-\bet H_1}
\end{equation}
is of trace class, \cf \cite[Lemma 9.5.3, p.~242]{cg:qgravity-book2}. Therefore, quantum statistics can be applied to this configuration.
\br\lar{4.6}
Note that the $n(\lam_i)$ are not uniformly bounded. Indeed, from \cite[equs. (9.4.60), (9.4.72), pp.~234--235]{cg:qgravity-book2}, we conclude
\begin{equation}
\lim_{i\ra \un}n(\lam_i)=\un.
\end{equation}
\er

Using this result, we combine the quantizations of the exterior and interior regions by defining
\begin{equation}\lae{4.28}
m_i=n(\lam_i),\qq\A\, i\in\N,
\end{equation}
so that the quantum-statistical results are identical in both regions, although the spatial eigenfunctions are, of course, different. We summarize this result in the following theorem.
\bt\lat{4.7}
Let $N=N^{n+1}$, $n=2m$, $m\ge 2$, be a Kerr-AdS black hole with all rotational parameters equal and with the metric defined in \fre{8.2.108} and \fre{8.2.15.1}. Assume that the negative cosmological constant $\Lam$ satisfies estimate \fre{3.18}. For any Cauchy hypersurface $\so=S_t$ in the exterior region, the exterior region can be canonically quantized so that the hyperbolic equation \re{3.1} in the quantum spacetime \re{3.2} admits a sequence of solutions that are products of temporal eigenfunctions and spatial tempered eigendistributions. These eigendistributions generate a Hilbert space $\mc H$ and may be regarded as mutually orthogonal unit vectors.

A similar result holds in the interior region of the black hole. Choosing the multiplicities $m_i$ of the spatial eigenvalues $\lam_i$ in the exterior region to equal the multiplicities $n(\lam_i)$ of the corresponding spatial eigenvalues in the interior region---which appears to be the only logical choice---makes the quantum-statistical results in the two regions identical. In particular, the respective partition functions, average energies, and entropies coincide.

Furthermore, because the spatial eigendistributions in the two regions form orthonormal bases of their respective Hilbert spaces, relation \re{4.28} can be used to define a unitary map between these Hilbert spaces under which the respective spatial Hamiltonians are unitarily equivalent because the corresponding eigenvalues coincide. Hence, the black hole information paradox does not exist at the quantum level.
\et

\section{Sasakian manifolds}\las{5}
We now prove that $\Ss[2m-1]$ is a Sasakian manifold using only the fact that the unit sphere has constant curvature $\ka=1$ and known results about $K$-contact manifolds. We also show that relation \fre{4.26.1} gives the scalar curvature of the slices $S_r$ in the interior region of the Kerr-AdS black holes considered here.

In \cite[p.~251]{sasaki:normal}, Sasaki and Hatakeyama defined \tit{normal} contact metric structures on odd-dimensional Riemannian manifolds $M=M^{2n+1}$, $n\ge 1$, equipped with a contact metric structure $(\F,\xi,\h,g)$ for which the tensor field $N^i_{jk}$ vanishes. In Section 5 of the same paper, they proved that odd-dimensional unit spheres $\Ss[2n+1]$ have this property. These spaces were later called Sasakian manifolds.

We give a short proof that an odd-dimensional unit sphere $\Ss[2n+1]$ is Sasakian, using the equivalent characterization in \cite[Definition--Theorem 6, (1)]{boyer:sasakian}, which can be stated as follows.
\bt
A Riemannian manifold $(M,g)$, $\dim M=2n+1$, is called Sasakian if there exists a unit Killing vector field $\xi$ such that the tensor field $\F$ of type $(1,1)$, defined by $\F(X)=-\nabla_X\xi$, satisfies the condition
\begin{equation}\lae{5.1.1}
(\nabla _X\F)(Y)=g(X,Y)\xi-g(\xi,Y)X
\end{equation}
for any pair of vector fields $X$ and $Y$ on $M$.
\et
\bc\lac{5.2.1}
The standard unit sphere $\Ss[2m-1]$ with the metric in \fre{2.10}---more explicitly, with the metric
\begin{equation}
ds^2=\s_{ab}d\theta^ad\theta^b +\mu_i^2d\f_i^2,
\end{equation}
\cf \re{2.14.2}, has a unit Killing vector field $\xi$ and satisfies the condition \re{5.1.1}. Using the coordinates $(x^\al)=(\theta^a,\f^i)$, the vector field $\xi$ can be expressed as
\begin{equation}
\xi=(\xi^a;\xi^i)=(0,\ldots, 0;1,\dots,1).
\end{equation}
The dual form $\h=(\h_\al)$
\begin{equation}
\h_\al=g_{\al\bet}\xi^\bet
\end{equation}
has the components
\begin{equation}\lae{5.5.1} 
\h=(\h_a;\h_i)=(0,\ldots,0;\mu_i^2).
\end{equation}
\ec
\bp
The vector field $\xi$ has unit length. We first prove that it is Killing; equivalently, $\h$ must satisfy
\begin{equation} 
\h_{\al;\bet}+\h_{\bet;\al}=0,
\end{equation}
where a semicolon indicates covariant differentiation and a comma indicates partial differentiation. The metric $g_{\al\bet}$ splits into two blocks
\begin{equation}
(g_{\al\bet})=(\s_{ab};\h_i\de_{ij}),
\end{equation}
hence
\begin{equation}
(g^{\al\bet})=(\s^{ab};\mu_i^{-2}\de^{ij}).
\end{equation}
At an arbitrary point on $\Ss[m-1]$, choose the coordinates $\theta^a$ to be Riemannian normal coordinates with respect to $\s_{ab}$. We then compute
\begin{equation}
\h_{a;b}=\h_{a,b}=0=\h_{b;a},
\end{equation}
\begin{equation}
\h_{i;j}=0=\h_{j;i},
\end{equation}
and
\begin{equation}
\h_{a;i}=-\frac12 \h_{i,a}=-\h_{i;a}.
\end{equation}
Hence, $\xi$ is a Killing vector field.

From the definition of $\F$ we infer that the condition \re{5.1.1} is equivalent to
\begin{equation}\lae{5.12.1}
\xi^\al_{;\bet\ga}=\h_\bet\de^\al_\ga-g_{\bet\ga}\xi^\al.
\end{equation}
On the other hand, because $\xi$ is a Killing vector field, its second covariant derivatives satisfy
\begin{equation}
\xi^\al_{;\bet\ga}=R^\al_{\hp{\al}\bet\ga\de}\xi^\de
\end{equation}
in any Riemannian manifold, \cf \cite[equ. (71.1), p.~237]{eisenhart}. Because $\Ss[2m-1]$ has constant curvature $\ka=1$, this identity implies relation \re{5.12.1}.
\ep

We now prove that the scalar curvature $R$ of the spacelike slices $S_r$ in the interior region of the Kerr-AdS black hole satisfies equation \fre{4.26.1}.
\bl\lal{5.3.1}
The scalar curvature $R$ of the spacelike slices $S_r$ in the interior region of the Kerr-AdS black hole satisfies equation \fre{4.26.1}.
\el
\bp
As noted after equation \re{4.22.1}, the induced metric on the slice $S_r$ is identical to the metric in \fre{2.72.1}, which we write more explicitly as
\begin{equation}\lae{5.14.1}
\begin{aligned}
ds^2&=d\tau^2+ C\s_{ab} d\theta^a d\theta^b + C(\mu_i^2\de_{ij}+b^2\mu_i^2\mu_j^2)d\f^id\f^j\\
&=d\tau^2+\rho^2(\s_{ab} d\theta^a d\theta^b + \mu_i^2\de_{ij} d\f^id\f^j+b^2\mu_i^2\mu_j^2d\f^id\f^j)\\
&=d\tau^2+\rho^2(g_{\al\bet}+b^2\h_\al\h_\bet)d x^\al d x^\bet,
\end{aligned}
\end{equation}
where $g_{\al\bet}$ is the metric on $\Ss[2m-1]$, \cf the notation in \rc{5.2.1}. Thus, the nonflat part of the metric is a positive constant multiple of a special deformation of the metric on the Sasakian manifold $\Ss[2m-1]$. The deformation is produced by the summand
\begin{equation}
b^2 \h_\al\h_\bet,
\end{equation}
in view of \re{5.5.1}. Crucially, $\h$ is the contact form of the Sasakian manifold. We may therefore apply Tanno's results in \cite[Section 2, p.~702]{tanno-important}. There, a D-homothetic deformation $g\ra {}^*g$ is defined by
\begin{equation}\lae{5.16.1}
{}^*g_{ij}=\al g_{ij}+\bet \h_i\h_j
\end{equation}
for constants $\al,\bet$ satisfying $\al>0$ and $\al+\bet>0$, where $g_{ij}$ is a Sasakian metric, \cf \cite[equ.~(2.11), p.~702]{tanno-important}. The deformation considered here belongs to this class with $\al=1$ and $\bet=b^2$. Tanno considered a manifold of dimension $2n+1$, and the scalar curvature of the deformed metric is
\begin{equation}
{}^*R=\al^{-1}R_0-\al^{-2}\bet 2n,
\end{equation}
where $R_0$ is the scalar curvature of the Sasakian manifold, \cf \cite[equ. (2.16)]{tanno-important}. In our case, the dimension is
\begin{equation}
2m-1=2(m-1)+1,
\end{equation}
i.e., $n=m-1$, which completes the proof of the lemma. 
\ep

\section{Quantum information and its unitary correspondence}\las{6}
Let $U$ denote the unitary map defined in \frt{4.7}
\begin{equation}\lae{6.1}
U: \mc {H_{\tup{int}}}\longrightarrow \mc {H_{\tup{ext}}}.
\end{equation}
This map sends an eigenbasis of the spatial Hamiltonian $H_{1,\tup{int}}$ onto an eigenbasis of the spatial Hamiltonian $H_{1,\tup{ext}}$, preserving the eigenvalues and their multiplicities. These eigenfunctions generate the respective Hilbert spaces. The eigenfunctions, or eigendistributions, $v_{ij}$ of $H_{1,\tup{ext}}$ are physically relevant because they are the spatial factors in solutions of the hyperbolic equation \fre{1.1}, which is the final result of the canonical quantization of the exterior region of the black hole. The Hamilton constraint, the diffeomorphism constraint, and the invariance of Riemannian metrics under coordinate changes have already been incorporated into the quantization process, \cf \cite[Chap.~4.2, p.~108]{cg:qgravity-book2}. Hence, $\mc {H_{\tup{ext}}}$ is the physical one-particle Hilbert space, and no further constraints need to be considered. The same applies to $\mc {H_{\tup{int}}}$.

Regard $\mc H_{\tup{ext}}$ as a real vector space
\begin{equation}
 \mc H_{\tup{ext},\R[]},
\end{equation}
and define the symplectic form
\begin{equation}
 \s_{\tup{ext}}(f,g)=2\imc\spd fg_{\tup{ext}}.
\end{equation}
Because $\mc H_{\tup{ext}}$ is already the physical
one-particle Hilbert space, the physical Weyl algebra may be defined
directly by
\begin{equation}
 \mf{W}_{\tup{ext}}=\mf{W}\left(\mc H_{\tup{ext},\R[]},\s_{\tup{ext}}\right).
\end{equation}
Its generators satisfy
\begin{equation}
 W_{\tup{ext}}(f) W_{\tup{ext}}(g)=\exp\left(-\frac{i}{2}\s_{\tup{ext}}(f,g)\right)W_{\tup{ext}}(f+g),
\end{equation}
\begin{equation}
 W_{\tup{ext}}(f)^{*}=W_{\tup{ext}}(-f),
\end{equation}
and
\begin{equation}
 W_{\tup{ext}}(0)=I.
\end{equation}

The symmetric Fock space is 
\begin{equation}
 \mc F_{\tup{ext}}=\mc F\left(\mc H_{\tup{ext}}\right).
\end{equation}
The canonical Fock representation is
\begin{equation}
 \pi_{\tup{ext}}: \mf W_{\tup{ext}}\longrightarrow\mc B\left(\mc F_{\tup{ext}}\right).
\end{equation}
The represented physical CCR algebra is
\begin{equation}
 \mf A_{\tup{ext}}=\pi_{\tup{ext}}\left(\mf W_{\tup{ext}}\right).
\end{equation}

To include the full von Neumann algebra of bounded observables generated by the represented Weyl operators, one passes to the von Neumann closure
\begin{equation}
 \mf M_{\tup{ext}}=\mf A_{\tup{ext}}''.
\end{equation}
The bounded observables are then the self-adjoint elements of $\mf M_{\tup{ext}}$.
The same construction yields
\begin{equation}
\mf W_{\tup{int}},\qq\mf A_{\tup{int}},\qq\mf M_{\tup{int}},
\end{equation}
from $\mc H_{\tup{int}}$. For a detailed overview of the Weyl CCR algebra, normal states, and the von Neumann closure we refer to \cite[Sections 5, 6, and 11]{derezinski} and \cite[Chapters 2.1, 2.3, and 2.4]{robinson:book1}

In view of \re{6.1} and the definition of $U$,
\begin{equation}
 H_{1,\tup{ext}}=U H_{1,\tup{int}}U^{*}.
\end{equation}

Since $U$ is unitary,
\begin{equation}
\s_{\tup{ext}}(Uf,Ug)=\s_{\tup{int}}(f,g).
\end{equation}
It therefore induces a unique CCR-algebra isomorphism
\begin{equation}
 \alpha_U: \mf W_{\tup{int}} \longrightarrow \mf W_{\tup{ext}},
\end{equation}
defined by
\begin{equation}
 \alpha_U\left( W_{\tup{int}}(f)\right) = W_{\tup{ext}}(Uf).
\end{equation}
Let
\begin{equation}
 \wh U= \Clk(U):\mc F_{\tup{int}} \longrightarrow\mc F_{\tup{ext}},
\end{equation}
be the second-quantized unitary.  Then
\begin{equation}
 \wh U \pi_{\tup{int}} \left(W_{\tup{int}}(f)\right)\wh U^{*} = \pi_{\tup{ext}}
 \left(W_{\tup{ext}}(Uf)\right).
\end{equation}
Consequently,
\begin{equation}\lae{6.19}
 \mf A_{\tup{ext}} =\wh U \mf A_{\tup{int}} \wh U^{*},
\end{equation}
and
\begin{equation}
 \mf M_{\tup{ext}} = \wh U \mf M_{\tup{int}} \wh U^{*}.
\end{equation}

Let $\rho_{\tup{int}}$ be the density operator representing any normal interior state. Define
\begin{equation}
 \rho_{\tup{ext}}=\wh U \rho_{\tup{int}}\wh U^{*}.
\end{equation}
For every interior observable
\begin{equation}
 A\in\mf M_{\tup{int}},
\end{equation}
define
\begin{equation}
 A_{\tup{ext}} = \wh U A\wh U^{*}.
\end{equation}
Then
\begin{equation}
 \tr\left(\rho_{\tup{ext}}A_{\tup{ext}}\right) =\tr\left(\rho_{\tup{int}}A\right).
\end{equation}
For any finite family $A_1,\ldots,A_k$,
\begin{equation}
 \tr\left(
 \rho_{\tup{ext}}
 A_{1,\tup{ext}}\cdots A_{k,\tup{ext}}\right) = \tr\left(\rho_{\tup{int}}A_1\cdots A_k\right).
\end{equation}
All spectral probabilities are preserved as well.

Let $\mc R$ be any reference system, \cf \cite[Section~3.3, p.~20]{preskill:2018chapter3}, and let
$\rho_{\mc R,\tup{int}}$ be a joint density operator. Then
\begin{equation}
 \rho_{\mc R,\tup{ext}}=\left(I_{\mc R}\otimes\wh U\right)\rho_{\mc R,\tup{int}}
 \left(I_{\mc R}\otimes\wh U^{*}\right).
\end{equation}
Hence, all entanglement and correlations with external systems are preserved. 

Thus we have proved:
\bpp
The unitary equivalence preserves the complete quantum-statistical
information carried by every normal state under the correspondence between the interior and exterior regions of the black hole.
\epp

\br
The equality of canonical partition functions, average energies, and
von Neumann entropies in \frt{4.7} is a consequence of this stronger
unitary equivalence.  It is not the complete content of the
information-preservation statement.
\er

\section{Conclusion}\las{7} 
We quantized the exterior region of an odd-dimensional Kerr-AdS black hole in which all rotational parameters are equal. The resulting hyperbolic equation
\begin{equation}
H_0\tilde u_{ij}-H_1 \tilde u_{ij}=0
\end{equation}
in the quantum spacetime
\begin{equation}
Q=(0,\un)\times \so
\end{equation}
has a sequence of solutions of the form
\begin{equation}
\tilde u_{ij}=w_i v_{ij},\qq 1\le j\le m_i,
\end{equation}
where the $w_i$ are eigenfunctions of the temporal Hamiltonian $H_0$ and the $v_{ij}$ are eigendistributions of the spatial Hamiltonian $H_1$. Each corresponding eigenvalue $\lam_i$ has multiplicity one for $H_0$ and multiplicity $m_i$ for $H_1$. For fixed $i$, the multiplicity $m_i$ is theoretically unbounded; no physical or mathematical obstruction imposes an upper bound. Because the von Neumann entropy and average energy tend to infinity as $m_i\ra\un$, \cf \frl{4.4}, the only logical and physically meaningful choice is to set $m_i$ equal to the corresponding multiplicity $n(\lam_i)$ obtained by maximizing the multiplicities in the interior quantization of the Kerr-AdS black hole, \cf \cite[Chapter 8]{cg:qgravity-book2}. As in the interior case, the $v_{ij}$ may be regarded as mutually orthogonal eigenvectors generating a complex Hilbert space. Thus, setting $m_i=n(\lam_i)$ implicitly defines a unitary map between the respective Hilbert spaces, and the respective Hamiltonians are unitarily equivalent because their spectra, including multiplicities, coincide. Consequently, as proved in the preceding section, the complete quantum-statistical information carried by normal states is preserved under the unitary correspondence between the interior and exterior regions; hence, there is no information paradox, \cf \cite{hawking:information,hawking:information2,page:information,page:information2,almheiri:entropy}.

\bt
The spatial eigenfunctions $v_{ij}=v_{ij}(\tau,x)$ may be regarded as gravitational radiation modes that emanate from the event horizon and decay exponentially at infinity, satisfying
\begin{equation}
v_{ij}(0,x)=0
\end{equation}
and
\begin{equation}\lae{6.10}
\lim_{\tau\ra \un} \abs{v_{ij}(\tau,\cdot)}_{m,M_0}=0\qq\A\,m\in \N,
\end{equation}
where $\abs{\cdot}_{m,M_0}$ denotes the norm in $C^m(M_0)$. Furthermore, $v'_{ij}$ also decays exponentially at infinity, and therefore
\begin{equation}\lae{6.11}
\sup_{x\in M_0}\int_0^\un(\abs{v_{ij}(\tau,x)}^2+\abs{v'_{ij}(\tau,x)}^2)\,d\tau<\un.
\end{equation}
\et
\bp
Recall the definition of $\tau$ in \fre{2.22} and estimate \re{2.27.2} on the same page
\begin{equation}
r^{-1}(\tau)\le c_2 e^{-c_1\tau}.
\end{equation}
Also recall from \frt{3.1} that
\begin{equation}
v_{ij}(\tau,x)=\rho^{-\frac{n-1}2}(\tau) u_{ij}(\tau)\psi_{ij}(x),
\end{equation}
where the $\psi_{ij}$ are the smooth spherical harmonics appearing in equation \fre{2.17}. The eigendistributions $u_{ij}$ have the initial values
\begin{equation}
u_{ij}(0)=0
\end{equation}
and
\begin{equation}
u_{ij}'(0)=1,
\end{equation}
and satisfy the estimate
\begin{equation}
\abs{u_{ij}(\tau)}+\abs{u_{ij}'(\tau)}\le c\qq\A\,\tau\in [0,\un),
\end{equation}
\cf \frt{2.13} and \frc{2.8}.
The proof of the claims, apart from the assertion concerning $v'_{ij}$ and inequality \re{6.11}, is now straightforward because $r(0)=r_2$, the radius of the event horizon.

\cvm
\nd 
To prove the asymptotic behavior of $v'_{ij}$ and inequality \re{6.11}, we differentiate $v_{ij}$ with respect to $\tau$
\begin{equation}
\begin{aligned}
v'_{ij}&=-\frac{n-1}2 \rho^{-\frac{n-1}2} \rho^{-1}\rho'u_{ij}\psi_{ij}+\rho^{-\frac{n-1}2} u'_{ij}\psi_{ij}\\
&=-\frac{n-1}2 \rho^{-\frac{n-1}2}\frac{r^2}{r^2+a^2} r^{-1}h^\frac12 u_{ij}\psi_{ij}+\rho^{-\frac{n-1}2} u'_{ij}\psi_{ij},
\end{aligned}
\end{equation}
where we used relation \fre{2.23}. Moreover, from the definition of $h$ in \re{2.20.1}, we deduce
\begin{equation}
r^{-1} h^\frac12\le c\qq\A\, r\ge r_2
\end{equation}
with a uniform constant $c$. Since $u_{ij}$, $u'_{ij}$, and $\psi_{ij}$ are uniformly bounded, the exponential decay of $v'_{ij}$ and inequality \re{6.11} follow from \fre{2.27.2}.
\ep

\br
Hawking radiation is obtained by quantizing a field on a prescribed classical black-hole background, \cf \cite{hawking:bh}. The gravitational radiation modes $v_{ij}$ in \frt{4.7} are the spatial parts of the eigenfunctions of the hyperbolic equation \fre{1.1}, which is the final result of the canonical quantization of the exterior region of the black hole. The Hamilton constraint, the diffeomorphism constraint, and the invariance of Riemannian metrics under coordinate changes have already been incorporated into the quantization process, \cf \cite[Chap.~4.2, p.~108]{cg:qgravity-book2}. Hence, $\mc {H_{\tup{ext}}}$ is the physical one-particle Hilbert space, and no further constraints need to be considered. The same applies to $\mc {H_{\tup{int}}}$. An elementary graviton is already fixed in the derivation of equation \re{1.1}, and its eigenvalue is included in the definition of the temporal Hamiltonian $H_0$.

Thus, the exterior gravitational radiation belongs to the quantized geometry itself. It is not an auxiliary test field and is not derived from a linearization of the classical Einstein equations about a fixed
black-hole metric.

This distinction also explains why the existence of the exterior radiation modes need not be interpreted as black-hole evaporation. The unitary interior-to-exterior radiation equivalence is a statement about
the quantum state and observable structures.  It does not require a time-dependent mass, a shrinking horizon, or a time-dependent inverse temperature.
\er
\providecommand{\href}[2]{#2}
\providecommand{\bysame}{\leavevmode\hbox to3em{\hrulefill}\thinspace}
\providecommand{\href}[2]{#2}



\end{document}